# About influence of differential rotation in convection zone of gaseous or fluid giant Planet (Uranus) onto the parameters of orbits of satellites.

**Sergey Ershkov***, Affiliation[1]: Plekhanov Russian University of Economics, Scopus number 60030998, e-mail: sergej-ershkov@yandex.ru

Affiliation[2]: Sternberg Astronomical Institute, M.V. Lomonosov's Moscow State University, 13 Universitetskij prospect, Moscow 119992, Russia,

**Dmytro Leshchenko**, Odessa State Academy of Civil Engineering and Architecture,Odessa, Ukraine, e-mail: leshchenko_d@ukr.net

**Elbaz I. Abouelmagd**, Celestial Mechanics and Space Dynamics Research Group (CMSDRG), Astronomy Department, National Research Institute of Astronomy and Geophysics (NRIAG), Helwan, Cairo 11421, Egypt,

e-mail: elbaz.abouelmagd@nriag.sci.eg

**Abstract**

Tidal interactions between Planet and its satellites are known to be the main phenomena, which are determining the orbital evolution of the satellites.

We suggest in the current research to take into consideration the additional well-known effect of differential rotation which obviously takes place in the gaseous or fluid convection zone of primary giant Planet (indeed, the aforementioned effect exists even not depending on the orbital evolution of satellites around host Planet).

Nevertheless, estimations for the contribution of the aforementioned effect of differential rotation in the Uranus system (including all its most massive satellites) let us exclude using such effect from calculations of mutual evolution of the eccentricity $e$ along with the semi-major axis $a$ for all satellites of Uranus (Planet of "ice" type).

It means that the Uranus can be considered in the analytic exploration of governing equations as to be the appropriate candidate for applying the modern ansatz [Efroimsky, 2015] (tidal dissipation effect depending on the tidal-flexure frequency $\chi$) in regard to estimations of eccentricity $e$ along with semi-major axis $a$ for satellites of Uranus.

We can see from the results of calculations in Section 3 that the *combined* system of governing equations (in the sense of *combined contributions* to the tidal dissipation from Uranus + from satellite) yields the really observed magnitudes of decelerations for semi-major axises of all the satellites of Uranus. Meanwhile, internal heat generation effects in the satellites (due to tidal dissipation effect) are much more than those which definitely take place in the Uranus, excepting the case of Ariel.

## 1. Introduction, the additional effect of differential rotation.

The main motivation of current research is the analytical exploration of tidal evolution of satellite's orbit when it moves onto quasi-elliptic trajectory around the primary *gaseous* or fluid Planet. In our previous research [1], we have obtained the governing equations for orbits of satellite which basically have been determined via appropriate tidal effects for the case of satellite's quasi-periodic rotation around (preferably) *rigid* Planet.

According to [1], here and below we define that the tidal effects are introduced by means of the Love number $k_2$ (associated with the response of potential of the deformed Planet in regard to the gravitational effect of the additional tidal bulge on its surface), as well as by the quality factor $Q$, which is inversely proportional with respect to the amount of energy dissipated by tidal friction; so, tidal effects are introduced in the combination $k_2/Q$ for Planet and satellite. Besides, we should note that, according to the Kepler's law of orbital motion, the square of mean motion:

$$n^2 = \frac{G(M+m)}{a^3}, \qquad (*)$$

here $m$ is the mass of satellite, $M$ is the mass of Planet (with equatorial radius $R$), $n$ is the osculating mean motion, $G$ is gravitational constant of the Newton's law of universal gravitation, $a$ is the semi-major axis.

Meanwhile, it is worth to note that theory of tidal effects is physically quite self-consistent in case of satellite's motion around the *rigid* Planet (also we should note that the aforesaid theory have been developed enough during last decades by a lot of authors in celestial mechanics). According to our understanding, the most natural ansatz has been developed in the comprehensive articles [2-3], where authors generalize their researches for exploring the tidal dissipation effect depending on the tidal-flexure frequency $\chi$. Namely, in this case the quality factor $Q$ of the Planet can be assumed depending on the tidal-flexure frequency $\chi$:

$$Q \propto \chi^{\alpha}, \quad \alpha = 0.16 \div 0.4$$

where frequency $\chi$ is apparently depending on the mean motion (*): $\chi = 2\,|\omega_P - n|$ ($\boldsymbol{\omega}_P$ being planet's spin rate).

It is very important to create the adequate physical model along with the mathematical model of the aforementioned influencing of tidal phenomena on the evolution of orbits of satellites with the main aim of the proper clarifying the results of data of astrometric observations.

That's why we should take also into consideration the additional well-known effect of differential rotation which obviously takes place in the gaseous or fluid convection zone of primary Planet [4-5] (indeed, the aforementioned effect exists even not depending on the orbital evolution of satellites around host Planet).

E.g., according to the data of helio-seismology [4], the Solar convection zone (i.e., the region of Sun with active convection-diffusion processes) is located at the outer part of Sun, $r > 0,6 \div 0,7 R$ (Fig.1), $r$ is the distance from the center of Sun:

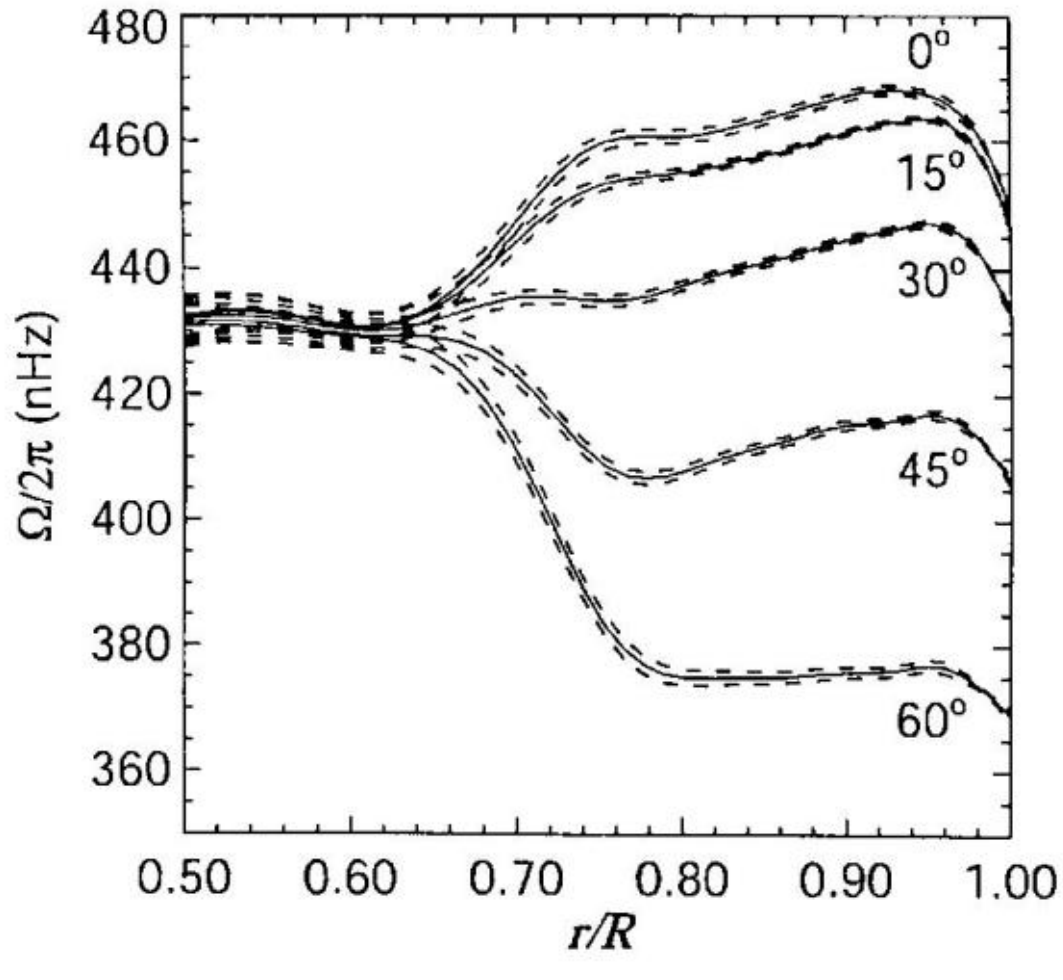

Fig.1. Distribution of Solar plasma angular velocities,

depending on radius of Solar zone [4-5].

The effect of differential rotation means that rotation of plasma on the surface of Sun is not uniform [5] (Fig.2):

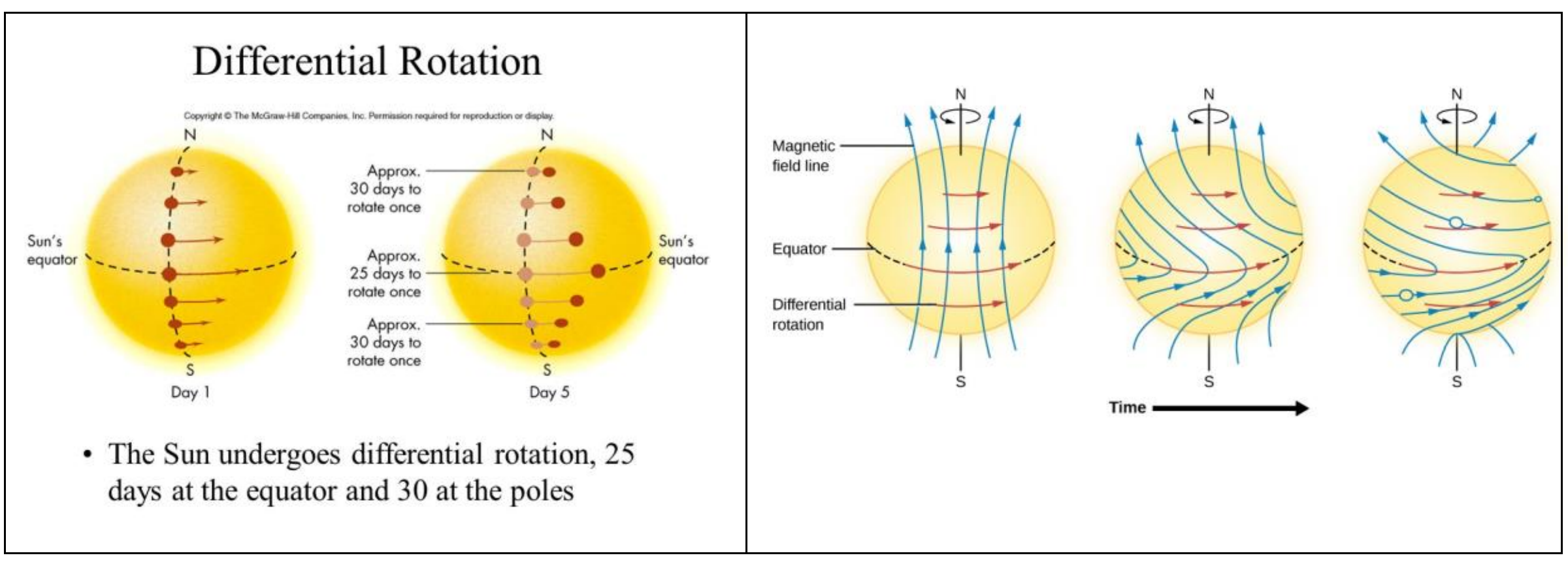

Fig.2. Schematic mechanism of the effect of differential rotation in Solar convection zone [4-5].

Thus, combining tidal effects along with the effect of differential rotation (for primary *gaseous* or fluid giant Planet), we should note that tidal bulge is additionally moving by differential rotation during its motion on the distorted surface of convection zone of the *gaseous* or fluid giant Planet.

It means that satellite is to be additionally shifted onto the angle $\delta\theta$ on its orbit by the aforementioned tidal bulge (according to additional displacement of bulge by the mechanism of differential rotation on the surface of Planet), whereas the total sum of angular shifting of satellite should be determined as below

$$\frac{1}{Q} \pm \delta\theta \qquad (1)$$

where $(1/Q)$ corresponds to the angular displacement by tidal effects, as first approximation (associated with the main *lmpq* tidal mode [2-3], influencing onto the tidal bulge: *lmpq = 2200*), but choosing the sign (±) of additional displacement onto the angle $\delta\theta$ depends on the acceleration or deceleration case of differential rotation's phenomenon near equator of the Planet (we should note that we know at least 10% of stars where differential rotation decelerates equatorial flows with respect to the polar flows on their surfaces [4]).

Moreover, we can estimate the aforementioned angle of displacement $\delta\theta$ for orbital displacements of satellites which have been taking place during tidal time-lag $\Delta t$ (overall time-lag is inverse proportionally to the tidal-flexure frequency $\chi$), via simple geometric consideration of displacement ($\delta\omega_P \cdot \Delta t$) of tidal bulge on equator with respect to the appropriate triangles "satellite - new position of tidal bulge (due to differential rotation) - center of mass of the Planet" ($R = const$):

$$\tan\Big((\delta\,\omega_p)\cdot\Delta t\Big) = \tan\big(\delta\,\theta\big)\cdot\left(\frac{\left|\vec{r}_s\right| - R}{R}\right), \qquad (2)$$

$$\Big\{\delta\,\theta << 1, \quad \Big((\delta\,\omega_p)\cdot\Delta t\Big) << 1\Big\} \qquad \Rightarrow$$

$$(\delta\,\omega_p)\cdot\Delta t = \delta\,\theta\cdot\left(\frac{\left|\vec{r}_s\right| - R}{R}\right), \qquad (3)$$

(here sign "*s*" denotes the case of satellite), where overall tidal time-lag $\Delta t$ [3] is:

$$\Delta t \;=\; \frac{2\pi}{\chi} \;=\; \frac{\pi}{\left|\omega_p - n\right|}, \qquad (4)$$

but variation of angular velocity $\delta\omega_P$ (due to differential rotation) could be estimated according to the results of [4] as below:

$$\delta\,\omega_p \sim (\omega_p)^{\beta}\,, \quad \beta = 0.15 \div 0.7 \qquad (5)$$

Let us note that formulae (2)-(3) along with (4)-(5) should be valid in all range of distances of satellite from the center of Planet, including the most distant case $r = a$; so, we obtain in this case (as first approximation by the Tailor series from (2)):

$$(\omega_p)^{\beta}\cdot\frac{\pi}{\left|\omega_p - n\right|} \;=\; \delta\,\theta\cdot\left(\frac{a - R}{R}\right), \qquad \left\{ n^2 = \frac{G(M+m)}{a^3} \right\} \quad \Rightarrow$$

$$\delta\,\theta \;=\; (\omega_p)^{(\beta-1)}\cdot\left(\frac{\pi\,R}{(a-R)\cdot\left|1 - \dfrac{\sqrt{G(M+m)}}{\omega_p\cdot a^{3/2}}\right|}\right), \qquad (6)$$

(for the definiteness, we will consider the case $\omega_P > n$ in the expression (6), so we should assume $|\omega_P - n| = \omega_P - n$). All variable functions in (1)-(6) depend on the quality factor $Q$, in general case (the tidal time lag should depend on the dissipation also).

## 2. The governing Eqns. for evolution of eccentricity, semi-major axis.

Taking into consideration the aforementioned angle of displacement $\delta\theta$ (6) for orbital displacements of satellites, we should update the systems of equations ($A$1), ($A$2) which have been presented in [1] for mutual evolution of the eccentricity $e$ along with the semi-major axis $a$ of the moons of primary Planet.

The secular evolution of the orbit and rotation are due to the dissipation of the energy of tides that leads to a geometrical lag between the tidal bulge and the line of centers and to a net torque. As such, both the tidal angle and net torque must depend on the friction and the associated energy dissipation rate. As we formulated in previous section, all variable functions in (1)-(6) depend on the quality factor $Q$, in general case.

Namely, we should use (1) instead of the expression for $(1/Q)$ in Eqns. ($A$1), ($A$2). In particular, we obtain for the tides raised in the primary Planet (as first approximation, we consider here the main *lmpq* tidal mode [2-3], *lmpq = 2200*):

$$\frac{da}{dt} = \frac{3k_2 mnR^5}{Ma^4}\left(1+\frac{51}{4}e^2\right)\cdot\left(\frac{1}{Q} \pm \delta\theta\right),$$

$$\frac{de}{dt} = \frac{57k_2 mn}{8M}\left(\frac{R}{a}\right)^5 e\cdot\left(\frac{1}{Q} \pm \delta\theta\right), \qquad (7)$$

where expression for $\delta\theta$ is given in (6).

But we also recall that we have (as first approximation) for the tides raised in the 1:1 spin–orbit satellite [1]:

$$\frac{da}{dt} = -\frac{21k_2^s MnR_s^5}{Q^s ma^4}e^2,$$

$$\frac{de}{dt} = -\frac{21k_2^s Mn}{2Q^s m}\left(\frac{R_s}{a}\right)^5 e \qquad (8)$$

Thus, the physically reasonable hypothesis should be assumed as below (*the mixed scenario*): - the tidal dissipation of the satellite in (8) is assumed to be equal to the *constant value*, but the tidal dissipation of the primary Planet should be assumed depending on tidal-flexure frequency $\chi$ and semi-major axis $a$ as suggested in (6)-(7).

## 3. Comparing the results for the combined system of equations (7)+(8).

Let us consider 1-st equation of the *combined* system of equations (7)+(8) in the sense of *combined contributions* to the tidal dissipation in case of Uranus (from Uranus + from satellite) as below, according to the assumption $e \to 0$. Besides, we should note that restriction $\delta\theta << (1/Q)$, which reduces accordingly the right parts of Eqns. (7), is valid for most massive satellites of Uranus (Planet of "ice" type):

$$\frac{da}{dt} = \frac{3k_2 mnR^5}{QMa^4}\left(1 + \frac{51}{4}e^2\right) - \frac{21k_2^s Mn \cdot (R_s)^5}{Q^s m}\left(\frac{1}{a}\right)^4 \cdot e^2,$$

$\Rightarrow$

$$\frac{da}{dt} \cong \frac{3k_2 mnR^5}{QMa^4} - \frac{21k_2^s Mn \cdot (R_s)^5}{Q^s m}\left(\frac{1}{a}\right)^4 \cdot e^2, \qquad (9)$$

where in fact we should neglect by $e^2$ contributions in (*da/dt*), and we should suggest in Eqn. (9) additionally the updating of the fundamental assumption in [2-3] for the quality factor $Q$ of the Planet depending on the tidal-flexure frequency $\chi$

$$Q = (E \cdot \chi)^{\alpha}, \quad (\alpha = 0.2 \div 0.4) \quad \chi = 2\left|\omega_p - n\right|, \qquad (10)$$

(here $E$ is the appropriate time-dimension *scale-factor*, which is supposed to be the proper constant for all satellites of Uranus). Besides, we should note that $k_2 = 0.104$, according to results of [6].

For all satellites of Uranus, we use the constants $(G{\cdot}M) \cong 5\ 793\ 939$ (in units of SI), $R =$

26200 km (radius of Uranus), $k_2$ = 0.104, $E$ = 646 596 770.7, $Q \cong 300$ in our calculations below (see [6]); besides, we choose for calculations the proper constants as pointed below

$$Q \;=\; (E\cdot\chi)^{\alpha},\;\; (\alpha = 0.25) \quad \chi = 2\left|\omega_p - n\right|,$$

$$\omega_p = (507.16^{\circ}/57.2958^{\circ})\, s^{-1}\,, \quad 1\, rad = 57.2958^{\circ}$$

So, we obtain from (9) (by using the actual data for calculations from the astrometric observations for Uranus, see [7]):

1) The case of satellite Ariel ($Q$ = 299.9):

| $G{\cdot}m$ | $n$ | $a$ | $e$ | $R_s$ | $da/dt$ | $k_2/Q$ | $k_2/Q_s$ |
|---|---|---|---|---|---|---|---|
| 90.3 | 2.49295234 | 190 929.79 | 0.00136551 | 577.9 | -0.3755 | -0.00034680 | 0.00003468 |

$$\Rightarrow \quad \left(\frac{k_2^s}{Q^s}\right)\cdot 10 \;=\; -\frac{k_2}{Q_{_plan}}$$

2) The case of satellite Umbriel ($Q$ = 311):

| $G{\cdot}m$ | $n$ | $a$ | $e$ | $R_s$ | $da/dt$ | $k_2/Q$ | $k_2/Q_s$ |
|---|---|---|---|---|---|---|---|
| 78.2 | 1.51614785 | 265 984.01 | 0.00424068 | 584.7 | -0.0712 | -0.00033444 | 0.03545093 |

$$\Rightarrow \quad \left(\frac{k_2^s}{Q^s}\right) = -106\cdot\frac{k_2}{Q_{_plan}}$$

3) The case of satellite Titania ($Q$ = 319.2):

| $G{\cdot}m$ | $n$ | $a$ | $e$ | $R_s$ | $da/dt$ | $k_2/Q$ | $k_2/Q_s$ |
|---|---|---|---|---|---|---|---|
| 235.3 | 0.72171832 | 436 281.94 | 0.00180148 | 788.9 | -1.5740 | -0.00032584 | 153.13158717 |

$$\Rightarrow \quad \left(\frac{k_2^s}{Q^s}\right) = -\,469\,960\cdot\frac{k_2}{Q_{_plan}}$$

4) The case of satellite Oberon ($Q$ = 321.7):

| $G{\cdot}m$ | $n$ | $a$ | $e$ | $R_s$ | $da/dt$ | $k_2/Q$ | $k_2/Q_s$ |
|---|---|---|---|---|---|---|---|
| 201.1 | 0.46669201 | 583 449.53 | 0.00140798 | 761.4 | -3.8484 | -0.00032330 | 3 111.96784699 |

$$\Rightarrow \left(\frac{k_2^s}{Q^s}\right) = -9625\,600 \cdot \frac{k_2}{Q_{_plan}}$$

5) The case of satellite Miranda ($Q$ = 273.1):

| $G{\cdot}m$ | $n$ | $a$ | $e$ | $R_s$ | $da/dt$ | $k_2/Q$ | $k_2/Q_s$ |
|---|---|---|---|---|---|---|---|
| 4.4 | 4.44519052 | 129 848.11 | 0.00132732 | 234.2 | -0.5174 | -0.00038081 | 0.65194540 |

$$\Rightarrow \left(\frac{k_2^s}{Q^s}\right) = -1712 \cdot \frac{k_2}{Q_{_plan}}$$

## 4. **Discussions & conclusion.**

Tidal interactions between Planet (Uranus) and its satellites are known to be the main phenomena, which are determining the orbital evolution of the satellites.

Meanwhile, the tidal dissipation model, which is used here to treat tidal friction with a power law of the tidal frequency for the tidal quality factor (10), is a model that has been derived for the dissipation of tides in solid bodies [2]. We refer the reader to the reviews by Ogilvie [8] and Mathis [9] for a detailed explanation showing the strong impact of the internal structure/rheology on tidal friction.

It is very important to create the adequate physical model along with the mathematical model of the aforementioned influencing of tidal phenomena on the evolution of orbits of satellites with the main aim of the proper clarifying the results of data of astrometric observations.

That's why we suggest in the current research to take also into consideration the additional well-known effect of differential rotation which obviously takes place in the gaseous or fluid convection zone of primary Planet [4-5] (indeed, aforementioned effect exists if even there is no satellite, which is orbiting around host Planet).

Nevertheless, estimations for the contribution of the aforementioned effect of differential rotation in the Uranus system (including all its most massive satellites) let us exclude using such effect from calculations in Eqn. (9) of mutual evolution of the eccentricity *e* along with the semi-major axis *a* for all satellites of Uranus (Planet of "ice" type), basing on the the data of astrometric observations.

It means that the Uranus can be considered in our analytic exploration of governing equations (7)-(8) as to be the appropriate candidate for applying the modern ansatz [3] for the sort of rheology which is good for solid planets (tidal dissipation effect depending on the tidal-flexure frequency $\chi$) in regard to estimations of evolution of the eccentricity *e* along with the semi-major axis *a* for satellites of Uranus.

So, as for the results in Section 3, we can conclude that such the modern ansatz [3] is self-consistent (in physical sense) to be applied to the analysis of the *combined* system of equations (9) in the sense of *combined contributions* to the tidal dissipation (from Uranus + from satellites).

Moreover, we can see from the results of calculations in Section 3 that the *combined* system of equations (9) (in the sense of *combined contributions* to the tidal dissipation from Uranus + from satellites) should yield the really observed values of decelerations for the semi-major axises of all the satellites of Uranus. Meanwhile, internal heat generation effects in the satellites (due to tidal dissipation effect) are much more than those which definitely take place in the Uranus, excepting the case of Ariel.

As for the range of the given magnitude of the quality factor $Q$ of the Planet (Uranus): we can see from data in Section 3 that quality factor is being varied in the range of magnitudes near $Q_0 \cong 300$ for each satellite: 1) ($Q_0$ - 0.03%) for Ariel, 2) ($Q_0$ + 3.67%) for Umbriel, 3) ($Q_0$ + 6.4%) for Titania, 4) ($Q_0$ + 7.23%) for Oberon, 5) ($Q_0$ - 8.97%) for Miranda. Using aforesaid magnitudes, we should obtain real observed values of accelerations for the semi-major axises for all the satellites of Uranus. If we begin our calculations from the given magnitude of the quality factor $Q_0 \cong 300$, we will obtain deviations in the calculated values of accelerations for the semi-major axises for all the satellites of Uranus (with respect to those which were real observed according to the data of astrometric observations [7]). In this case, we should correct our model (9) by taking into account the additional effects, e.g. we should take into consideration the internal heat generation in the Uranus or we should take into account the effect of differential rotation as prescribed by Eqn. (7).

The last but not least, we should especially emphasize in conclusion that the presence (or absence) of the differential rotation is the only criterion to decide could the sort of rheology which is good for solid planets (tidal dissipation effect depending on the tidal-flexure frequency $\chi$) be applied, or could it be applied as for the gas or liquid objects. If planet rotates as rigid body (the negligible effect of differential rotation) we can choose the aforesaid sort of rheology for solid planets.

According to the data of astrometric observations, the negligible effect of differential rotation was detected earlier on the surface of Uranus. So, it can be considered as "solid-type" planet.

Also, some remarkable articles should be cited, which concern the problem under consideration, e.g. [10]-[11] (on the topic of differential rotation on Jupiter), [12]-[13] (on the topic of gravity field's distortion due to the large-scale gas or plasma flows inside the Saturn's and Jupiter's envelope), [14] (on the topic of internal flows inside the differentially rotating fluid spherical body with rigid envelope), and [15]-[18] (on the topic of tides in rotating fluids and on the topic of evolution of the star's structure, rotation and tidal dissipation in its external convective envelope with tidal inertial waves excited in the convection zone). There are several concept studies devoted to design space-based missions to Uranus (and Neptune) which should be cited [19]-[23].

The rotational state of Uranus, inparticular its spin angular momentum, could be measured or, at least, constrained dynamically (with aim to estimate its upper limit) by exploiting General Relativity as a tool: the Lense-Thirring effect governs the small-scaled orbital precessions induced by the spin angular momentum of the central body which may be, in principle, extracted from the motion of natural and/or artifical satellites as proposed for Jupiter and the spacecraft Juno in [24].

In addition, works [25] (regarding special regimes of flows induced on surface of primary by tidal effects described by Laplace tidal equation under the actual creeping approximation [26]) and [27] (regarding non-linear regimes of satellites rotations) should be mentioned accordingly as well.

## Acknowledgements

Authors are thankful to unknown esteemed Reviewer with respect to his valuable efforts and advice which have improved structure of the article significantly (especially for advice regarding the interconnexion between measurement of angular momentum of Uranus and the Lense-Thirring effect).

## Conflict of interest

The authors declare that there is no conflict of interests regarding the publication of this article.

Remark regarding contributions of authors as below:

In this research, Dr. Sergey Ershkov is responsible for the general ansatz and the solving procedure, simple algebra manipulations, calculations, results of the article in Sections 1-3 and also is responsible for the search of approximate solutions.

Prof. Dmytro Leshchenko and Prof. Elbaz I. Abouelmagd are responsible for theoretical investigations, whereas Prof. Dmytro Leshchenko is responsible for the deep survey in literature in Section 4 on the problem under consideration.

All authors agreed with results and conclusions of each other in Sections 1-4.